# Dynamic Competition of Negative Bias Temperature Instability Mechanisms and Anomalous Recovery Overshoot in $HfO_x/AlO_x$-Gated $InO_x$ FETs

**Zih-Hao Dai[1*], Kazuki Ishiyama[2,3], Xuanhedong Gao[1], Chia-Tsong Chen[2], Toshifumi Irisawa[2], Tatsuro Maeda[2], Mitsuru Takenaka[1] and Kasidit Toprasertpong[1]**

[1] Department of Electrical Engineering and Information Systems, The University of Tokyo, Bunkyo-ku, Tokyo, Japan
[2] Semiconductor Frontier Research Center, National Institute of Advanced Industrial Science and Technology, Tsukuba, Ibaraki, Japan
[3] Department of Electrical and Electronic Engineering, Nihon University, Chiyoda-ku, Tokyo, Japan

*Author to whom any correspondence should be addressed.

**E-mail:** zhdai@mosfet.t.u-tokyo.ac.jp

## Abstract

The complex mechanisms of negative bias temperature instability (NBTI) and recovery overshoot in ultrathin $InO_x$ field-effect transistors with $SiO_2$ and $HfO_x/AlO_x$ gate dielectrics are investigated. A $HfO_x/AlO_x$-gated device exhibits positive threshold-voltage shifts during NBTI, whereas a $SiO_2$-gated reference device, in which the dielectric nonideality is suppressed, exhibits negative shifts under the same channel electrostatic conditions. This comparison supports competing positive and negative contributions from dielectric electron trapping and channel defect charging, respectively. During recovery at zero gate bias, the threshold voltage of the $HfO_x/AlO_x$-gated devices moves beyond its initial value, producing a negative overshoot. Increased $HfO_x$ thickness substantially increases the positive shift during NBTI while negative overshoots remain comparable, further supporting the dielectric and channel contributions. The negative recovery overshoot can be explained by the coexistence of the dielectric charge trapping, which causes a positive shift but relaxes faster, and channel defect charging, which causes a persistent negative shift.

## 1. Introduction

Oxide semiconductors (OS) have attracted considerable attention as channel materials for monolithic three-dimensional (M3D) integration because they combine favorable electron transport properties, low off-state currents, and compatibility with low-temperature processing [1-5]. These characteristics enable the integration of additional transistor layers above existing circuitry within a limited thermal budget. In particular, the low off-state current of OS field-effect transistors (FETs) is attractive for capacitorless two-transistor zero-capacitor (2T0C) gain-cell dynamic random-access memory (DRAM), in which charge retention depends strongly on the leakage characteristics of the transistors [6-10]. Realizing these applications, however, requires stable electrical characteristics throughout repeated operation. Threshold-voltage ($V_{th}$) instability can alter transistor drive current and circuit switching conditions, reducing timing margins in logic circuits and compromising retention and readout margins in memory cells [11]. Understanding and controlling bias temperature instability (BTI) is therefore essential for reliable OS-based integrated circuits.

The mechanism of BTI in OS FETs is complicated by the coexistence of processes associated with the semiconductor channel, the gate dielectric, and their interface. In particular, negative bias temperature instability (NBTI) refers to changes in transistor characteristics under prolonged negative gate-to-source bias at a given temperature. This is particularly important for access transistors in memory cells during data retention, where a gate voltage is held at zero while a source potential is nonzero depending on the storing states. Depending on the device structure and stress conditions, changes in defect charge states and carrier trapping can produce threshold-voltage shifts of either polarity. Consequently, the polarity of the applied gate stress does not necessarily determine the direction of the threshold-voltage shift. Studies of OS transistors have reported anomalous bias-stress responses and proposed mechanisms involving oxygen-related defects [12-14], dielectric charge trapping [15-16], and hydrogen-related processes [17-18]. For devices incorporating high-k gate dielectrics, decoupling the relative impact of channel and dielectric is particularly important because their contributions may occur simultaneously, making it difficult to get a comprehensive picture of actual mechanisms. After removal of the stress bias, transistor characteristics may evolve toward their initial state, a process referred to as recovery. The recovery of BTI is also complicated when there are multiple BTI mechanisms [15, 17]. On the other hand, since dynamics of BTI recovery are also expected to have unique behaviors, carefully examining both the threshold-voltage shift during BTI and their recovery kinetics is expected to help distinguish the contribution of each BTI mechanism on the target devices under given stress conditions.

In this work, we investigate NBTI and recovery in ultrathin $InO_x$ FETs with $HfO_x/AlO_x$ and thermally grown $SiO_2$ gate dielectrics. The $HfO_x/AlO_x$ stack allows the combined effects of electron trapping in the dielectric or at the dielectric interface (mentioned here as dielectric electron trapping) and $InO_x$ channel-related instability to be examined, while $SiO_2$-gated devices serve as a reference for investigating solely the impact of channel-related instability while minimizing the dielectric electron trapping [19–20]. We observe an anomalous negative recovery overshoot in the $HfO_x/AlO_x$-gated devices: following a positive threshold-voltage shift during negative-bias stress, the threshold voltage shifts beyond its initial value during recovery at zero gate bias. Building on our preliminary report at SSDM2026 [21], the present work provides a more precise comparison of NBTI among devices with different gate dielectrics by ensuring similar channel electrostatic conditions during stress and examines the effects of $HfO_x$ thickness to investigate the channel and dielectric contributions to the NBTI and its recovery behaviors.

## 2. Experiment

Figure 1 illustrates the device structures and fabrication process flow of long-channel, back-gated $InO_x$ field-effect transistors (FETs) with $AlO_x$ passivation and two different gate-dielectric configurations: $HfO_x/AlO_x$ and $SiO_2$. For the $HfO_x/AlO_x$-gated devices, a Si substrate covered with a 100-nm-thick $SiO_2$ layer was first cleaned using a standard wet-cleaning procedure. A Ti/Pt back-gate electrode was then formed by electron-beam evaporation. Subsequently, a gate-dielectric stack consisting of 5-nm-thick $HfO_x$ and 1-nm-thick $AlO_x$, a 2.3-nm-thick $InO_x$ channel, and a 10-nm-thick $AlO_x$ passivation layer were sequentially deposited by plasma-enhanced atomic layer deposition (PEALD) at 200 °C. A device with the $HfO_x$ thickness of 9 nm was also prepared for the discussion in the last part. The 1-nm-thick $AlO_x$ interlayer between $HfO_x$ and $InO_x$ has been reported to act as a diffusion barrier that suppresses excess hydrogen incorporation from $HfO_x$ into the $InO_x$ channel [16, 22]. The devices were then annealed at 300 °C in an $N_2$ atmosphere. The active area was defined by wet etching, followed by selective etching of the $AlO_x$ passivation layer to expose the source/drain contact regions. Ni source/drain contact pads were subsequently formed by electron-beam evaporation.

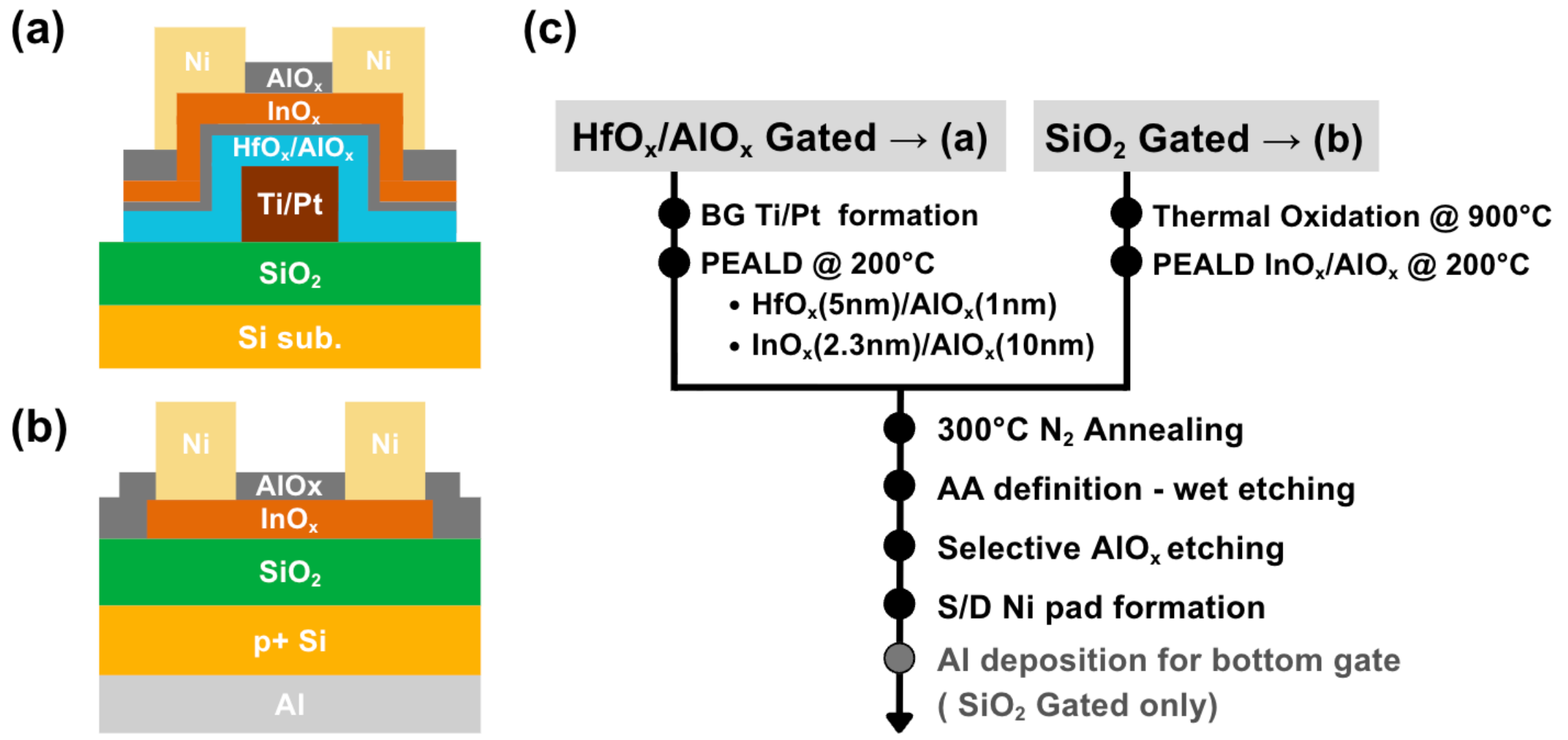

**Figure 1.** Schematic cross sections of back-gated $InO_x$-FETs with (a) $HfO_x/AlO_x$ and (b) $SiO_2$ gate dielectrics and (c) their fabrication process flow.

For the $SiO_2$-gated devices, a 16-nm-thick $SiO_2$ gate dielectric was grown on a heavily doped p-type Si substrate by thermal oxidation at 900 °C. The $InO_x$ channel and $AlO_x$ passivation layer, with thicknesses of 2.3 and 10 nm, respectively, were then sequentially deposited by PEALD at 200 °C. The subsequent annealing, active-area definition, contact-opening etch, and Ni contact formation followed the same process sequence as that used for the $HfO_x/AlO_x$-gated devices. An additional Al layer was deposited on the backside of the Si substrate to provide electrical contact to the p-Si substrate, which works as a back gate in this structure. Both device structures had a channel length of 100 µm and a channel width of 50 µm.

Figure 2 presents cross-sectional transmission electron microscopy (TEM) images of the two device structures. Both devices exhibit comparable $InO_x$ channel thicknesses of approximately 2.3

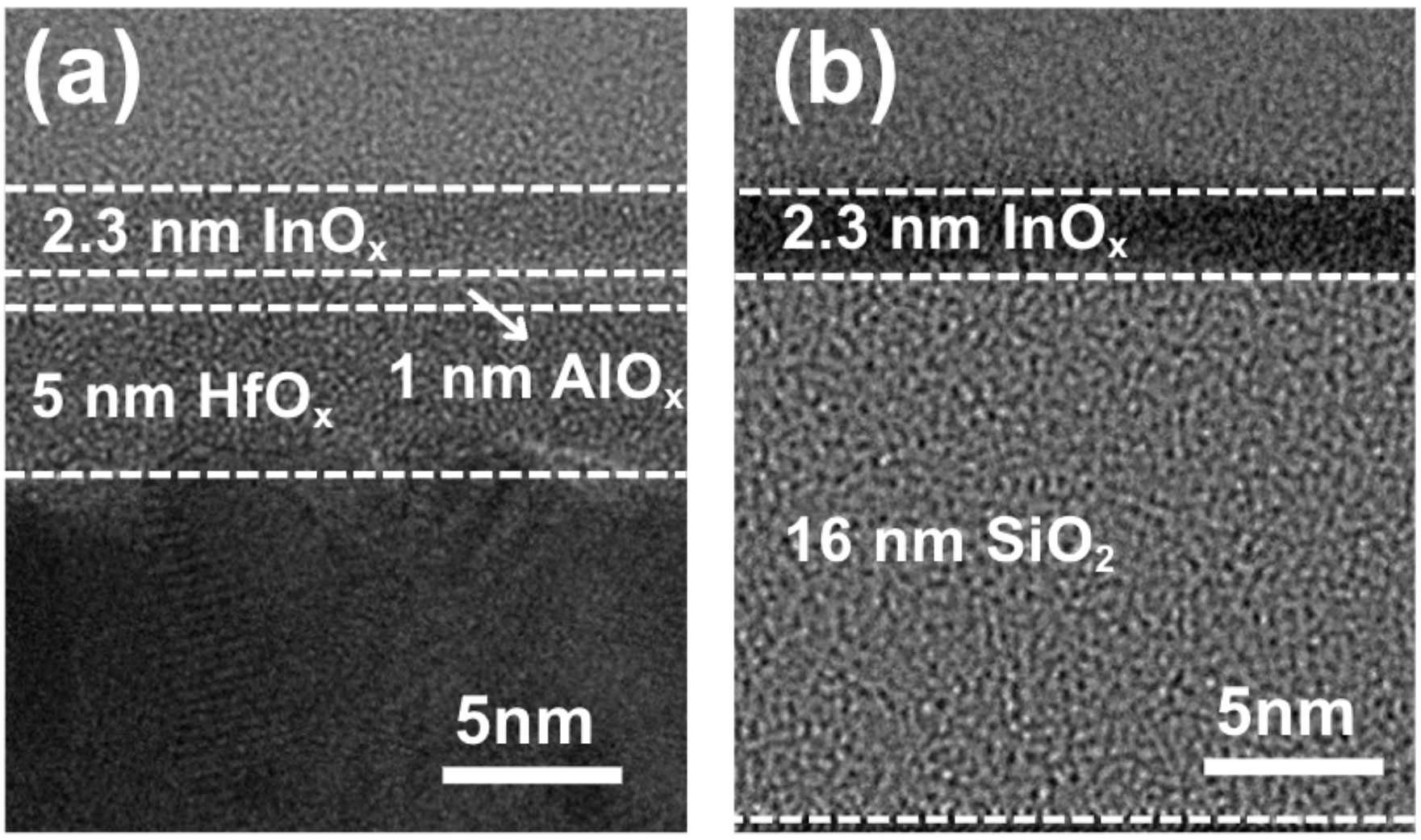

**Figure 2.** TEM images of $InO_x$-FETs with (a) $HfO_x/AlO_x$ and (b) $SiO_2$ dielectric layer.

± 0.1 nm. No discernible lattice fringes are observed in the $InO_x$ channels after annealing, confirming an amorphous channel structure.

Figure 3 illustrates the measure–stress–measure (MSM) protocol used to evaluate negative bias temperature instability (NBTI). Each measurement consisted of a 1-hour negative-bias stress phase followed by a 1-hour recovery phase. During stress, a negative gate voltage ($V_{g,\ stress}$) was applied between successive transfer-characteristic measurements. During recovery, the gate voltage was maintained at 0 V between measurements. The stress and recovery phases were periodically interrupted to acquire $I_D$–$V_G$ transfer characteristics using a drain–source voltage of $V_{DS}$ = 50 mV, with each sweep lasting 0.1 s. The delay between the end of each stress interval and the start of the subsequent $I_D$–$V_G$ sweep was set to 100 ns to minimize recovery before measurement. The threshold voltage ($V_{th}$) was extracted using the constant-current method. The initial threshold voltage ($V_{th,0}$) was obtained from the transfer curve measured immediately before the first stress interval, and the threshold-voltage shift was defined as $\Delta V_{th} = V_{th} - V_{th,0}$. The stress voltages were chosen to ensure similar channel potentials among different devices, as will be discussed subsequently.

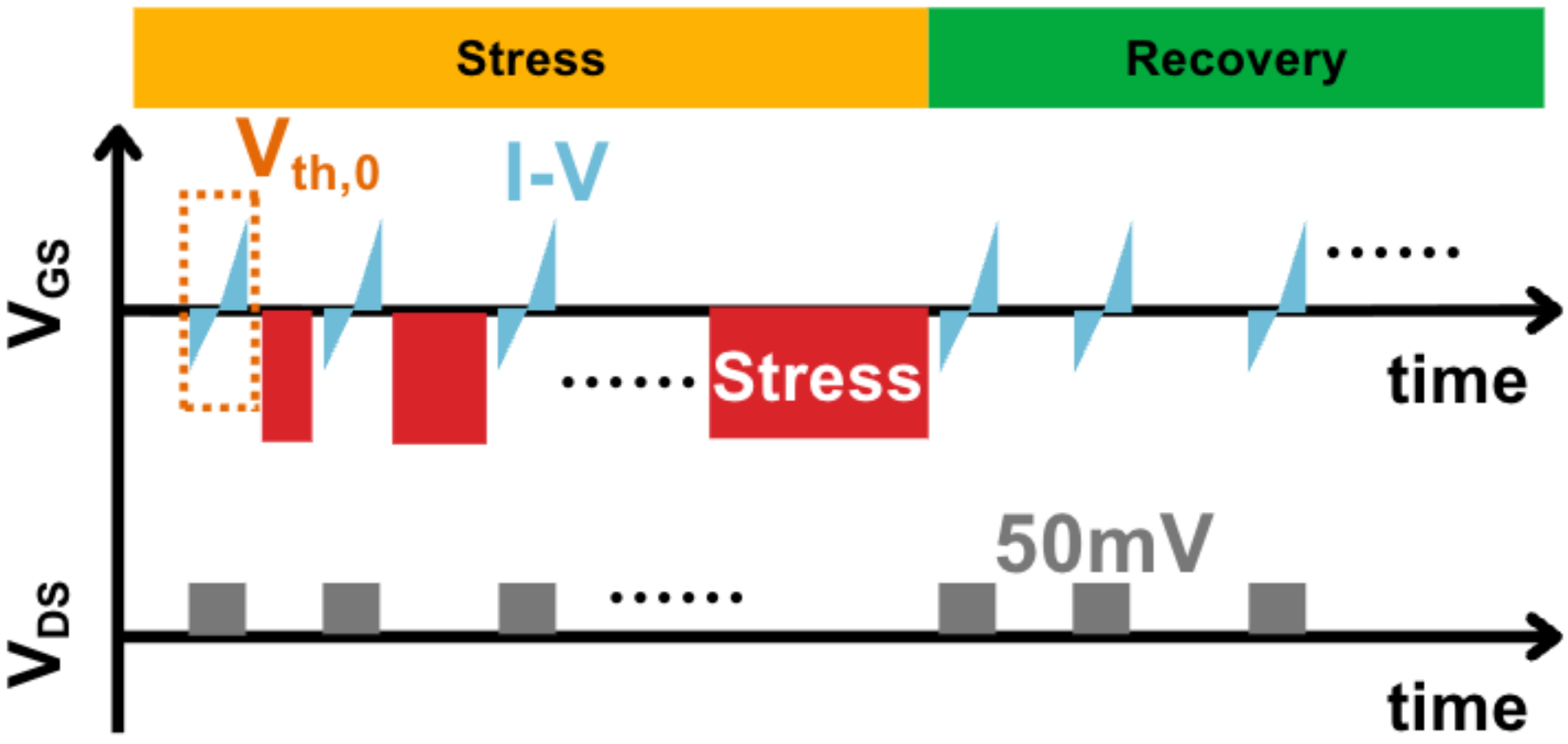

**Figure 3.** Schematic of the measure–stress–measure (MSM) protocol used for NBTI characterization, consisting of a 1 h negative-bias stress phase followed by a 1 h recovery phase at $V_g$ = 0 V. Transfer characteristics were periodically measured at $V_{DS}$ = 50 mV using a sweep duration of 0.1 s. The initial threshold voltage, $V_{th,0}$, was extracted from the transfer curve acquired before the first stress interval.

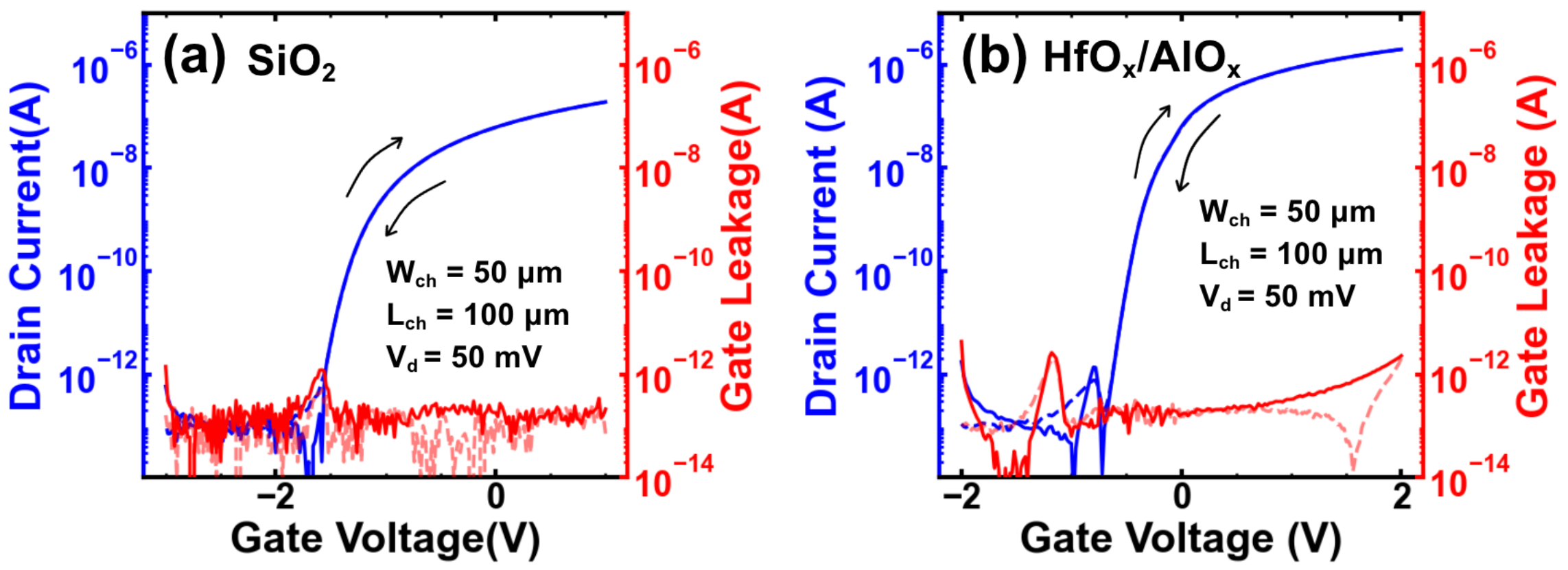

**Figure 4.** Transfer characteristic and gate leakage of $InO_x$-FET with (a) $SiO_2$ and (b) $HfO_x/AlO_x$ gate dielectric.

## 3. Results and Discussion

Figures 4 (a) and (b) show the initial transfer characteristics and gate leakage currents of the $SiO_2$- and $HfO_x/AlO_x$-gated $InO_x$ FETs, respectively. While both devices show negligible hysteresis and no significant leakage current, the $HfO_x/AlO_x$-gated device exhibits an improved minimum subthreshold swing (SS) of 76.3 mV/dec than the $SiO_2$-gated device (102.9 mV/dec) due to a better gate control of $HfO_x/AlO_x$.

To choose appropriate stress voltages for a fair comparison between the devices with different gate dielectrics during stress, one-dimensional electrostatic simulations of the MOS structures were performed by solving Poisson's equation with carrier densities calculated using Boltzmann statistics. The simulations were used to evaluate the out-of-plane band profiles, channel electric fields, and the position of the Fermi level relative to the conduction-band edge. Figure 5(a) shows the simulation path of the one-dimensional simulation. The simulation represents the central region of the long-channel devices and neglects lateral electrostatic variations near the source and drain contacts [23-24]. The permittivities, electron affinities, band gaps and gate work functions, used in the simulations are listed in Table 1. Identical channel thicknesses, donor concentrations, and dielectric constants were assumed for the two structures. The effective relative permittivities of $HfO_x$ and $AlO_x$ were estimated from C–V measurements using a two-layer series-capacitance model. A channel donor concentration of $\sim 10^{19}$ $cm^{-3}$ was assumed, a value commonly reported for ultrathin $InO_x$-based OS channels [25-27].

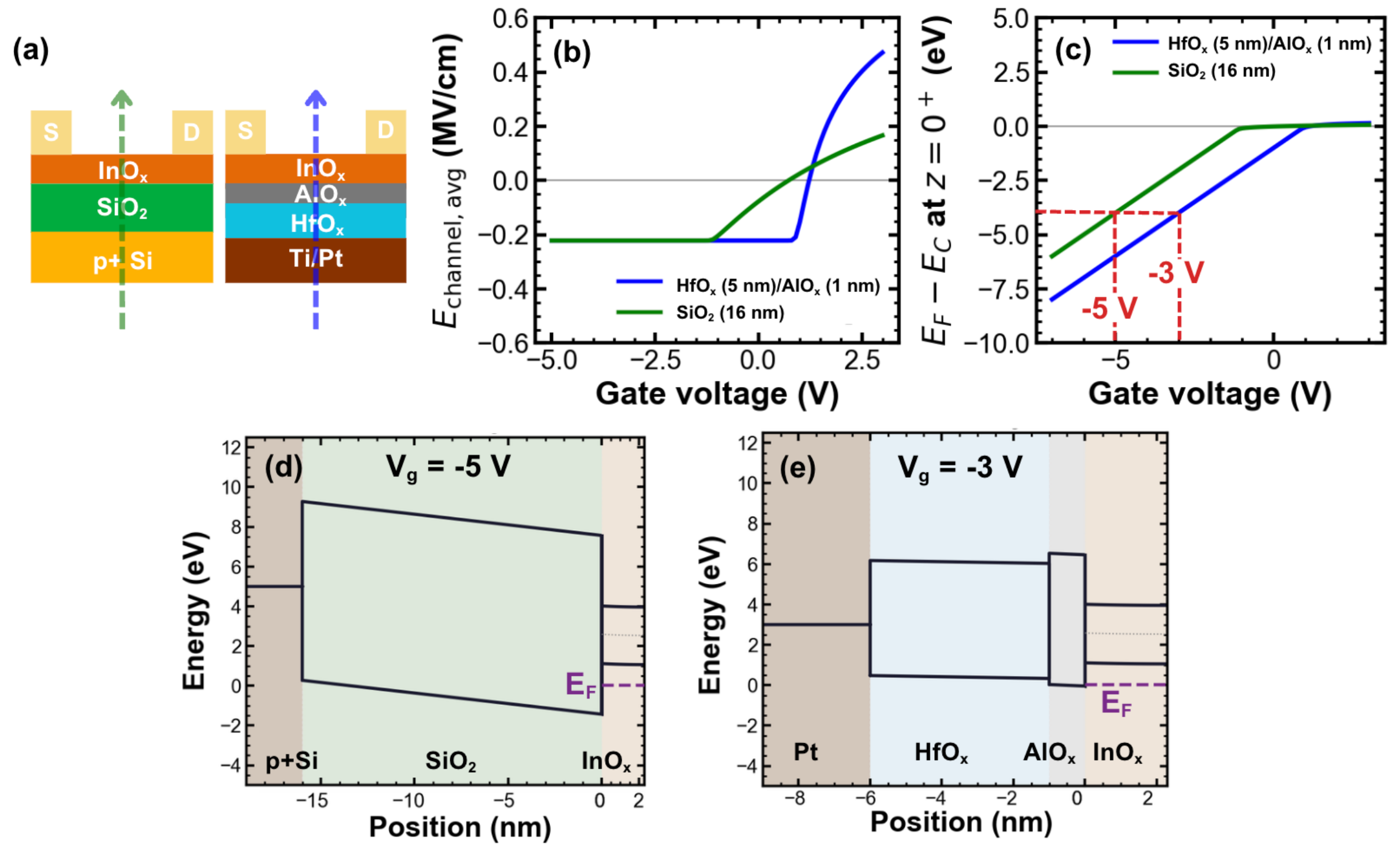

**Figure 5.** 1-D electrostatic simulations for selecting comparable negative-bias stress conditions in $SiO_2$- and $HfO_x/AlO_x$-gated $InO_x$ FETs. (a) Device schematics showing the vertical simulation paths (dashed arrows). Gate-voltage dependence of (b) the electric field $E_{channel,avg}$ averaged over the channel thickness and (c) the Fermi-level position relative to the conduction-band edge, $E_F - E_C$ in the $InO_x$ channel at the channel/dielectric interface ($z = 0^+$). Energy-band diagrams for (d) the $SiO_2$-gated device at $V_g = -5$ V and (e) the $HfO_x/AlO_x$-gated device at $V_g = -3$ V. These biases are selected in the NBTI tests to match the Fermi-level position and channel electric fields. Identical channel parameters are assumed for both structures.

Figure 5(b) shows the electric field $E_{channel,avg}$ averaged over the channel thickness under different gate voltages. As the gate voltage becomes increasingly negative, the $InO_x$ channel becomes fully depleted under the assumption that mobile minority holes are suppressed, and the average channel electric field $E_{channel,avg}$ approaches a saturation value given approximately by $E_{channel,avg} \approx -\frac{qN_D t_{os}}{2\varepsilon_{os}}$, where $q$ is the elementary charge, $N_D$ is the donor concentration, $t_{os}$ is the channel thickness, and $\varepsilon_{os}$ is the absolute permittivity of the channel. Under these assumptions, the same channel parameters yield the same saturated average field for both gate-dielectric structures.

Table 1. Parameters used in the one-dimensional electrostatic simulations [28-33].

| Parameter | $SiO_2$ | $HfO_x$ | $AlO_x$ | $InO_x$ |
|---|---|---|---|---|
| Relative permittivity, $\varepsilon_r$ | 3.9 | 14.5[a] | 5.6[a] | 8.9 |
| Electron affinity, χ (eV) | 0.9 | 2.5 | 2.0 | 4.45 |
| Bandgap, $E_g$ (eV) | 9.0 | 5.7 | 6.5 | 2.9 |

[a] Extracted from C–V measurements in this work.
The simulations were performed at 300 K with an $InO_x$ channel donor concentration of $N_D = 1 \times 10^{19}$ $cm^{-3}$. Gate work functions of 5.15 and 5.65 eV were used for $p^+$-Si and Pt, respectively.

Matching the channel field alone, however, does not establish equivalent channel conditions. Figure 5(c) therefore compares the Fermi-level position in the $InO_x$ channel relative to the conduction-band edge, $E_F - E_C$ at the channel/dielectric interface, as the Fermi level determines the defect ionization inside the channel. Although the channel field saturates after full depletion, $E_F - E_C$ continues to decrease as the gate voltage becomes more negative. The simulations indicate that stress voltages of approximately −5 V for the $SiO_2$-gated device and −3 V for the $HfO_x/AlO_x$-gated device provide similar $E_F - E_C$ values while both devices remain in the field-saturation regime. These stress voltages were selected to compare the NBTI responses of the two device structures under similar average channel electric fields and $E_F - E_C$ values. Figures 5(d) and 5(e) show the corresponding energy-band diagrams for the $SiO_2$- and $HfO_x/AlO_x$-gated structures, respectively, at these voltage conditions.

Based on the electrostatic comparison described above, gate stress voltages of −5 and −3 V were applied to the $SiO_2$- and $HfO_x/AlO_x$-gated devices, respectively. Figures 6(a)–6(b) show the evolution of the transfer characteristics of the $SiO_2$-gated $InO_x$ FETs under a gate stress voltage of −5 V at 25 and 85 °C, respectively. The $SiO_2$-gated device serves as a reference for investigating the instability and its possible origin in the $InO_x$ channel. During stress, the transfer curves shift toward more negative gate voltages, with a larger shift observed at higher temperatures. After 1 hour of stress, $\Delta V_{th}$ reaches −30.1 mV at 25 °C and −188.9 mV at 85 °C. This temperature dependence implies that the negative threshold-voltage shift is a thermally assisted process. One possible explanation is a change in the charge states of oxygen-related defects in the $InO_x$ channel. Under negative gate bias, a decrease in the Fermi-level position relative to the channel band edges can favor electron emission from occupied defect states [34]. Candidate processes include the ionization of oxygen vacancies and oxygen interstitials, represented as

$$V_O^0 \rightleftharpoons V_O^{2+} + 2e^- \tag{1}$$

$$O_i^{2-} \rightleftharpoons O_i^0 + 2e^- \tag{2}$$

Figures 6(c)–6(d) show the subsequent evolution of the transfer characteristics during recovery at $V_g$ = 0 V, where the transfer curves shift back toward the positive direction after removing the negative-voltage stress. Within the proposed oxygen-defect picture, this behavior is consistent with the reversal of stress-induced charge-state changes through electron recapture.

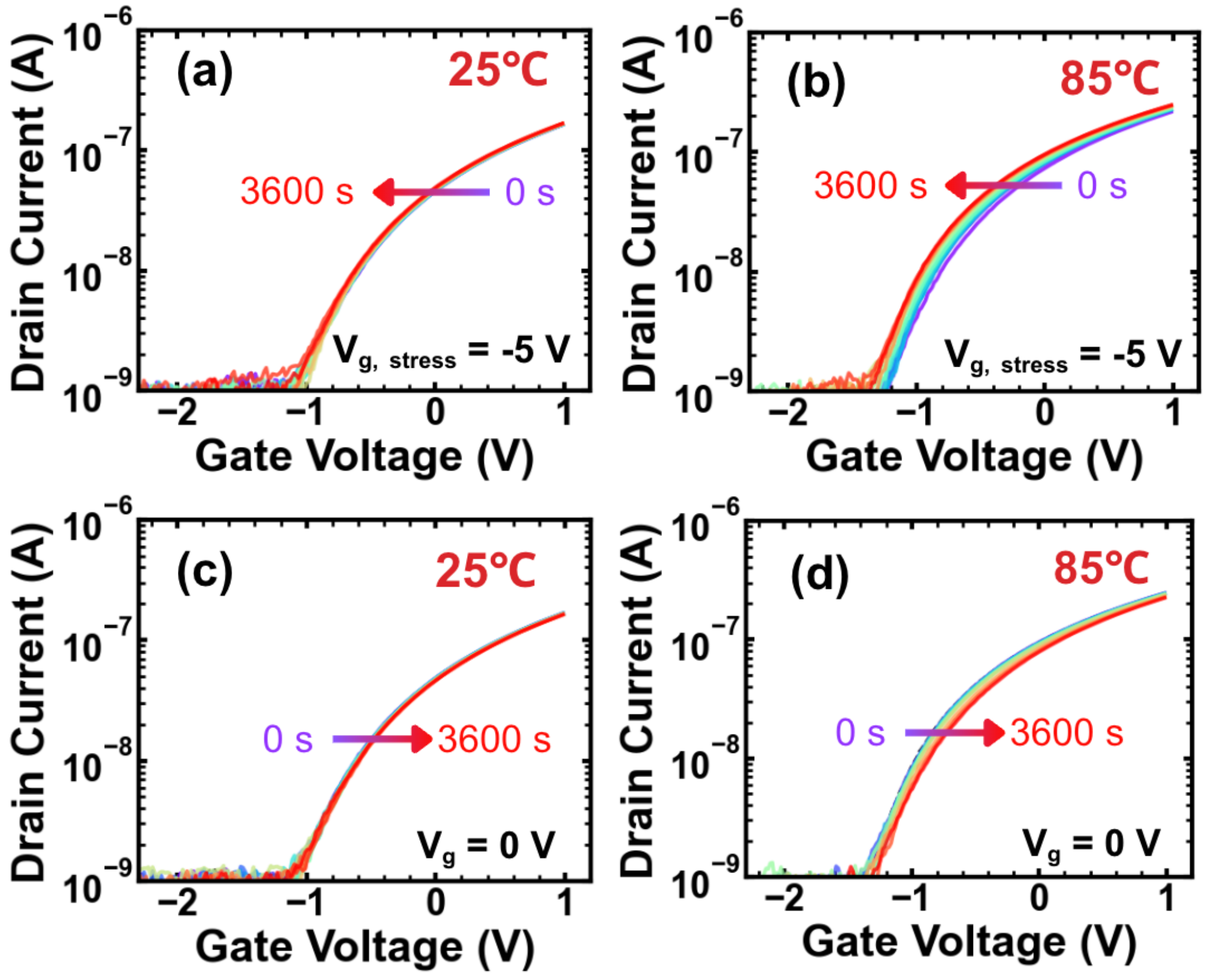

**Figure 6.** Evolution of transfer curves of $InO_x$-FET with $SiO_2$ dielectric during (a)-(b) NBTI and (c)-(d) recovery at (a)(c) 25°C and (b)(d) 85°C.

Figures 7(a)–7(c) show the evolution of the transfer characteristics of the $InO_x$ FETs with a 5-nm-thick $HfO_x$/1-nm-thick $AlO_x$ gate-dielectric stack under a gate stress voltage of −3 V. In contrast to the negative shifts observed in the $SiO_2$-gated device, the $HfO_x/AlO_x$-gated device exhibits little change at 25 °C and increasingly positive shifts at higher temperatures. Figure 7(d) summarizes the time evolution of $\Delta V_{th}$. The shifts after 1 hour of stress are approximately +3, +34, and +73 mV at 25, 65, and 85 °C, respectively. The positive shift is attributed to electron trapping in the gate-dielectric stack. Under negative gate bias, electrons injected from the gate electrode may become trapped in dielectric defects [15-16], producing a positive contribution to $\Delta V_{th}$. Here, we know from the result of $SiO_2$-gated device, where NBTI is expected to be affected only by the channel, that the charge state change in the channel is one of the origins of NBTI and it causes a negative threshold-voltage shift. Therefore, the threshold-voltage shift of the $HfO_x/AlO_x$-gated device during NBTI can be described by the combination of the two components: the impact from the $InO_x$ channel caused by charge state change that results in a negative threshold-voltage shift and the impact from the $HfO_x/AlO_x$ dielectric caused by gate-injected electron trapping that results in a positive threshold-voltage shift. Within this interpretation, the observed positive shift indicates that the positive contribution from gate-injected electron trapping in $HfO_x/AlO_x$ exceeds the contribution from the channel charge state during stress.

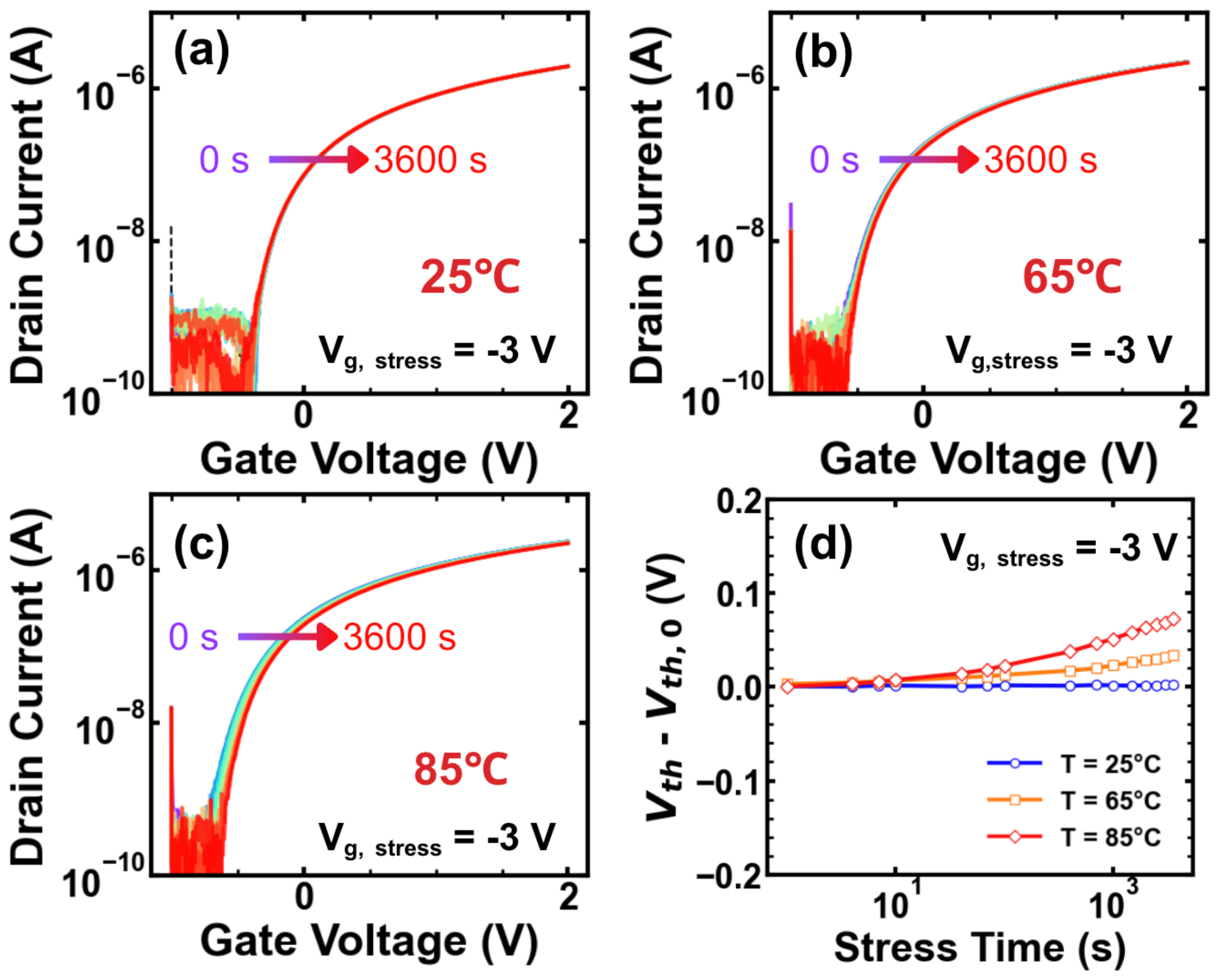

**Figure 7.** Evolution of (a)-(c) transfer curves at (a) 25°C, (b) 65°C, (c) 85°C, and (d) $V_{th}$ of $InO_x$-FET with $HfO_x/AlO_x$ dielectric during NBTI.

Figures 8(a)–8(c) show the transfer characteristics during subsequent recovery at $V_g$ = 0 V after 1 hour of stress, and Fig. 8(d) summarizes the corresponding evolution of $\Delta V_{th}$. Following the removal of the stress bias, $V_{th}$ shifts back in the negative direction. At elevated temperatures, particularly at 85°C, recovering $V_{th}$ moves even beyond its initial value, resulting in a negative recovery overshoot. This behavior is attributed to the two opposing contributions that recover on different effective time scales. The recovery of gate-injected electron trapping component occurs through electron detrapping from $HfO_x/AlO_x$, reducing the positive shift accumulated during stress, whereas the reversal of channel defect charging reduces the negative shift during stress. If the positive contribution decays more rapidly than the negative contribution, the latter becomes apparent during recovery, allowing the net $\Delta V_{th}$ to cross below zero during a given time scale of the recovery. The overshoot can therefore be interpreted as the emergence of a more persistent negative component that was partly masked by electron trapping in $HfO_x/AlO_x$ during stress. Electron trapping in $HfO_x/AlO_x$ dielectric is a fast component during both stress and recovery, resulting in net positive $\Delta V_{th}$ during NBTI and fast shifting back of $V_{th}$ to the negative direction.

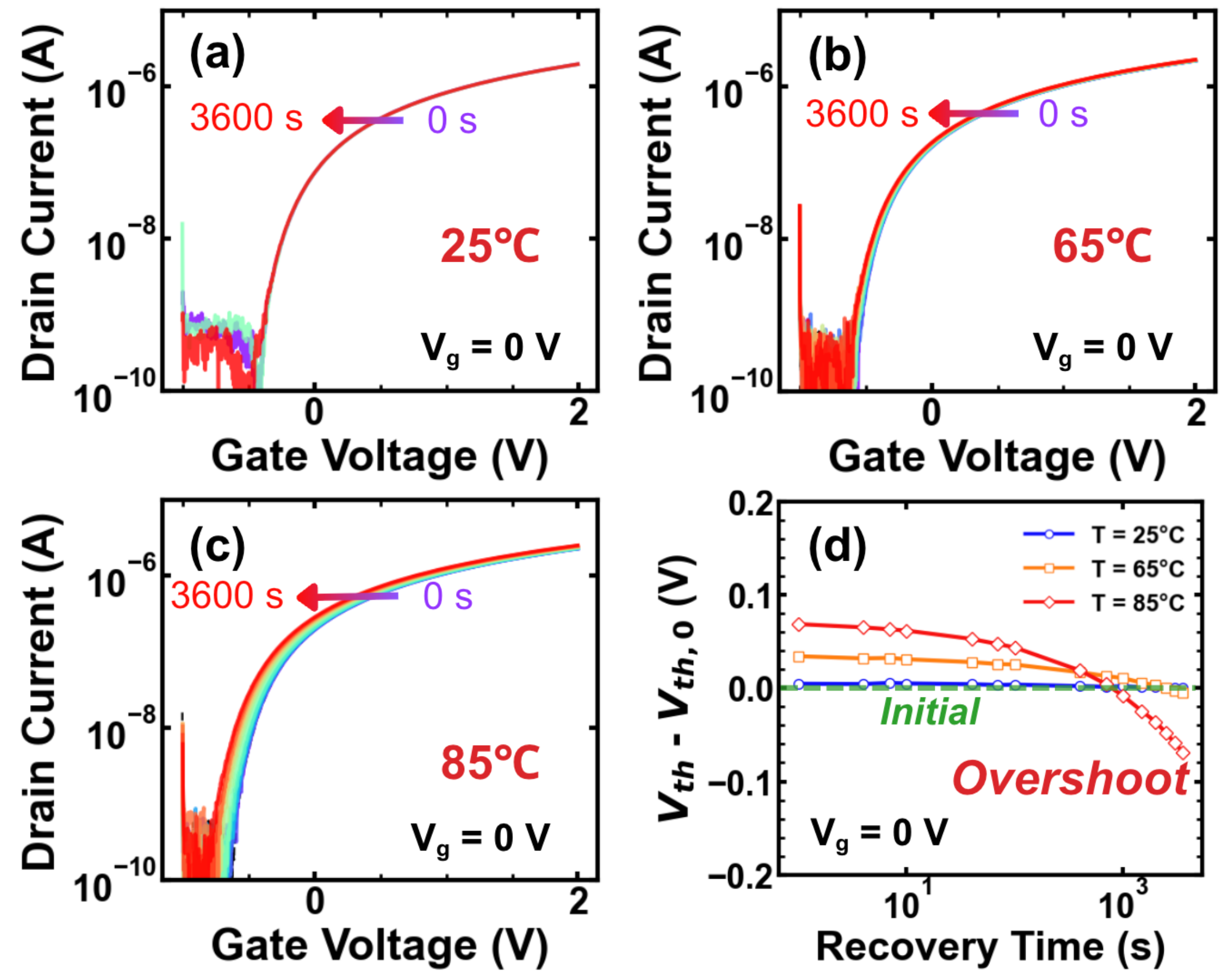

**Figure 8.** Evolution of (a)-(c) transfer curves at (a) 25°C, (b) 65°C, (c) 85°C, and (d) $V_{th}$ of $InO_x$-FET with $HfO_x/AlO_x$ dielectric during recovery at 0 V gate voltage.

A possible microscopic origin of this persistent component is the formation of metastable oxygen-related configurations in the channel. During NBTI, the change in the charge state of oxygen interstitials described by Eq. (2) may be accompanied by local structural rearrangement. In the excess oxygen peroxide model, an excess oxygen atom $(O_i^{2-})$ can form an $O-O$ bond with a neighboring oxygen atom upon electron removal $(O_i^0)$, producing a peroxide configuration $(O_2^{2-})$ [34-36]. First-principles calculations have shown that peroxide configurations can form spontaneously in certain local atomic environments [37]. After electron reinjection, these configurations can remain metastable because their reversal requires overcoming a structural energy barrier. Such a mechanism may contribute to the persistent negative component and apparent negative recovery overshoot observed in the present $InO_x$ devices. This is one possible explanation, although the electrical measurements alone do not uniquely identify the underlying defect configuration. Figure 9(a) compares the transfer characteristics of the $HfO_x/AlO_x$-gated device at 85 °C before stress, after 1 hour of stress, and after the subsequent 1-hour recovery period. The comparison clearly illustrates the positive shift during stress and the negative overshoot relative to the initial transfer curve after recovery. Figure 9(b) summarizes the corresponding threshold-voltage shifts as a function of temperature. The negative residual shift becomes more pronounced at higher temperatures. This trend reflects the combined temperature dependence of the processes occurring during stress and recovery.

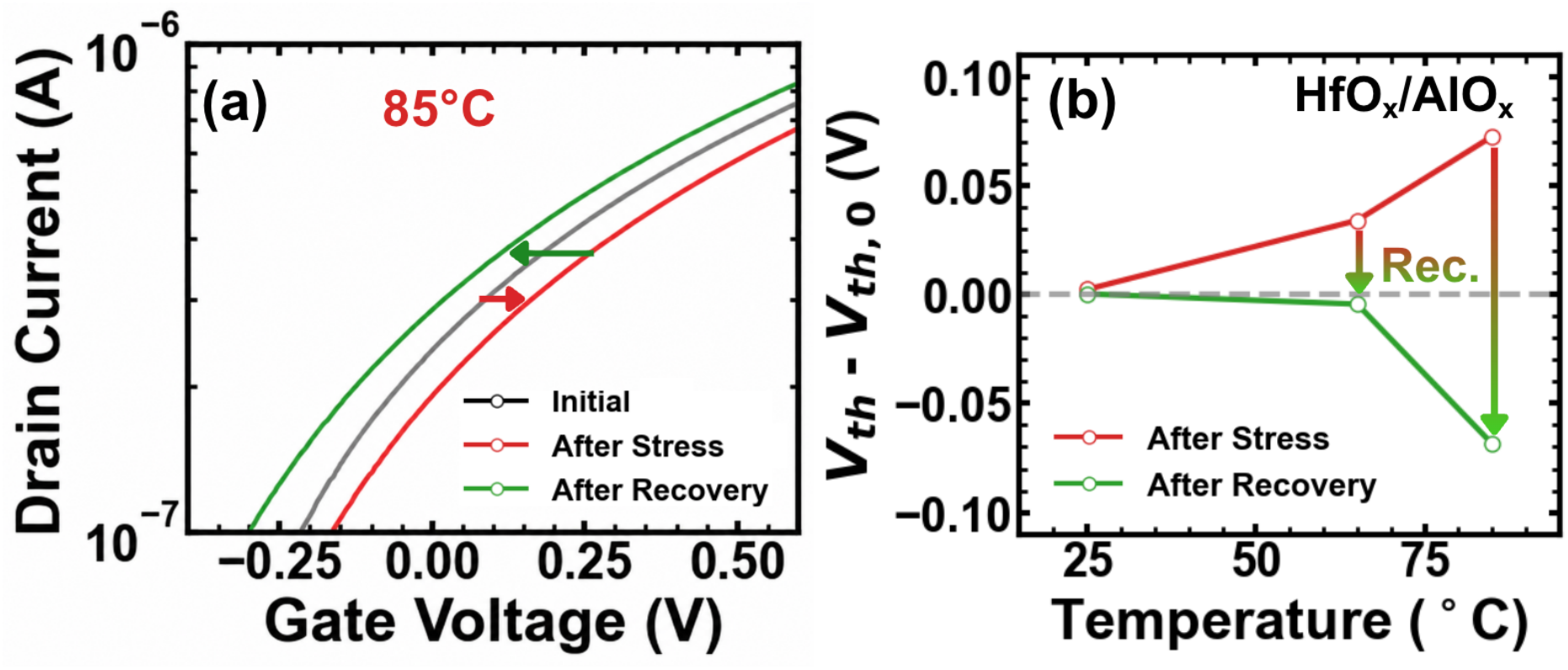

**Figure 9.** (a) Zoomed-in $I_d$-$V_g$ curves at 85°C and (b) temperature-dependent $V_{th}$ shift after 1-hour stress and 1-hour recovery for device with $HfO_x/AlO_x$ gate dielectric. The negative overshoot magnitude increases with temperature.

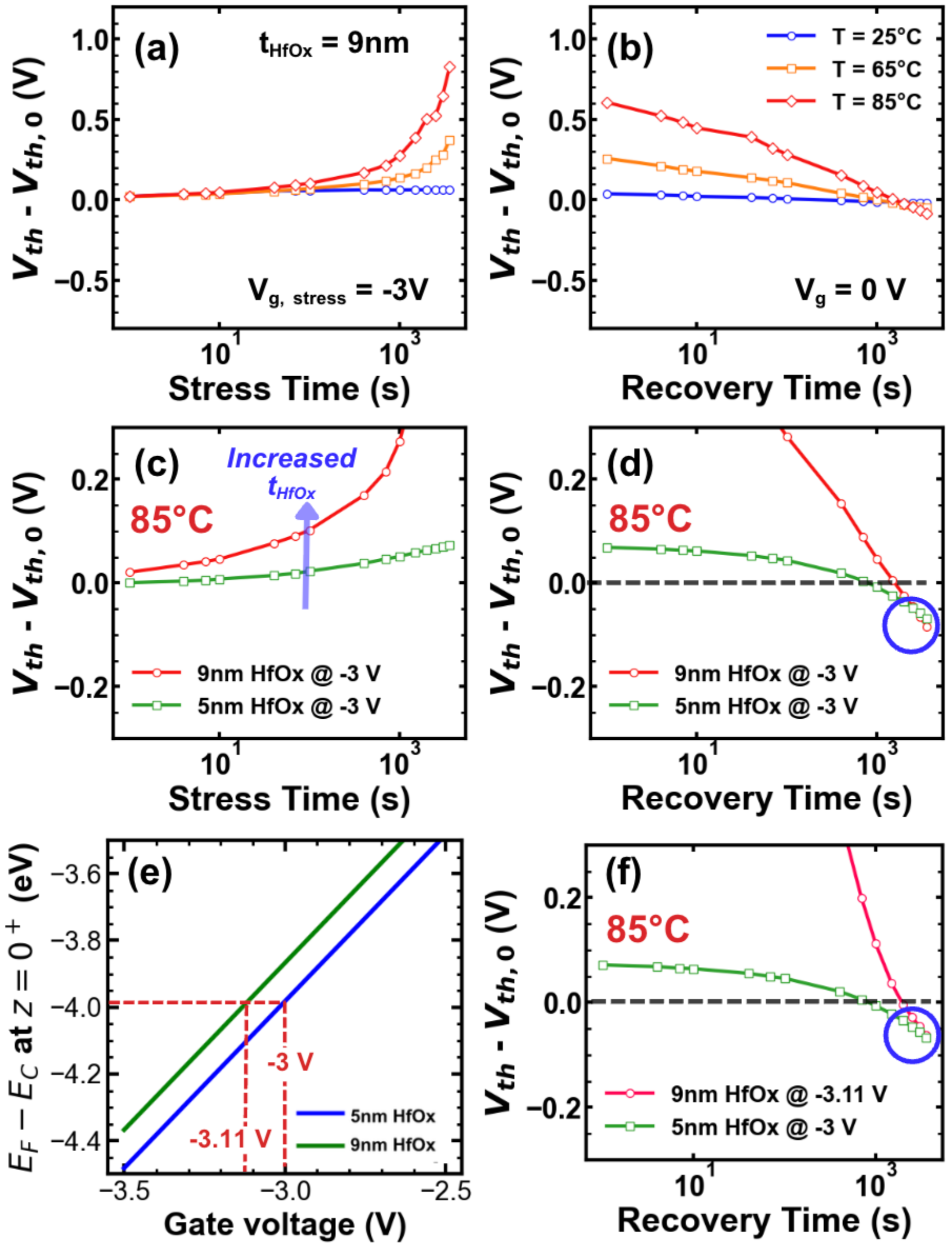

**Figure 10.** Effect of $HfO_x$ thickness on NBTI and recovery in $HfO_x/AlO_x$-gated $InO_x$ FETs. Time evolution of the $V_{th}$ shift for the device with 9-nm-thick $HfO_x$ during (a) stress at $V_g = -3$ V and (b) recovery at $V_g = 0$ V at different temperatures. Comparison between devices with 5- and 9-nm-thick $HfO_x$ at 85 °C during (c) stress at the same $V_g = -3$ V and (d) subsequent recovery. (e) Simulated gate-voltage dependence of $E_F - E_C$ at the channel-side gate-dielectric interface ($z = 0^+$), identifying stress voltages of −3 V and −3.11 V for the 5- and 9-nm devices, respectively, that yield the same $E_F$ position and comparable average channel electric fields. (f) Recovery at 85 °C following stress at these matched conditions. All recovery measurements were performed at $V_g = 0$ V. The thicker $HfO_x$ device exhibits a larger positive shift during stress, while comparable negative recovery overshoots are observed under both comparison conditions, further supporting a channel-related contribution to the overshoot.

To investigate the influence of gate-dielectric thickness, an additional device with a 9-nm-thick $HfO_x$ layer was examined. Figures 10(a) and 10(b) show its threshold-voltage evolution during stress at $V_g = -3$ V and subsequent recovery at $V_g = 0$ V. The positive shifts after 1 hour of stress are 63.7, 370.5, and 828.1 mV at 25, 65, and 85 °C, respectively, substantially larger than those of the device with 5-nm-thick $HfO_x$. For a comparable volumetric density of bulk defects, a thicker $HfO_x$ layer contains more potential trapping sites per unit area, which may contribute to the larger positive threshold-voltage shift observed during stress. This interpretation is consistent with previous reports of thickness-dependent electron trapping in $HfO_x$ gate dielectrics [38-39]. The lower gate-dielectric capacitance of the thicker stack may also contribute to the larger voltage shift, since a given effective trapped-charge density produces a larger threshold-voltage shift when the capacitance is smaller [38].

Figures 10(c) and 10(d) directly compare the 5- and 9-nm devices at 85 °C under the same stress voltage of −3 V. Despite the substantial difference in the positive stress-induced shift, the negative overshooting shifts remaining after 1 h of recovery are comparable. To calibrate the channel potential condition, the position of $E_F$ was also simulated, as shown in Fig. 10(e). A stress voltage of −3.11 V for the 9-nm device was selected to match the Fermi-level position with that of the 5-nm device stressed at −3 V. Figure 10(f) shows that comparable negative recovery overshoots are obtained under these matched conditions, supporting our model that the negative recovery overshoot is attributed to a channel-related contribution.

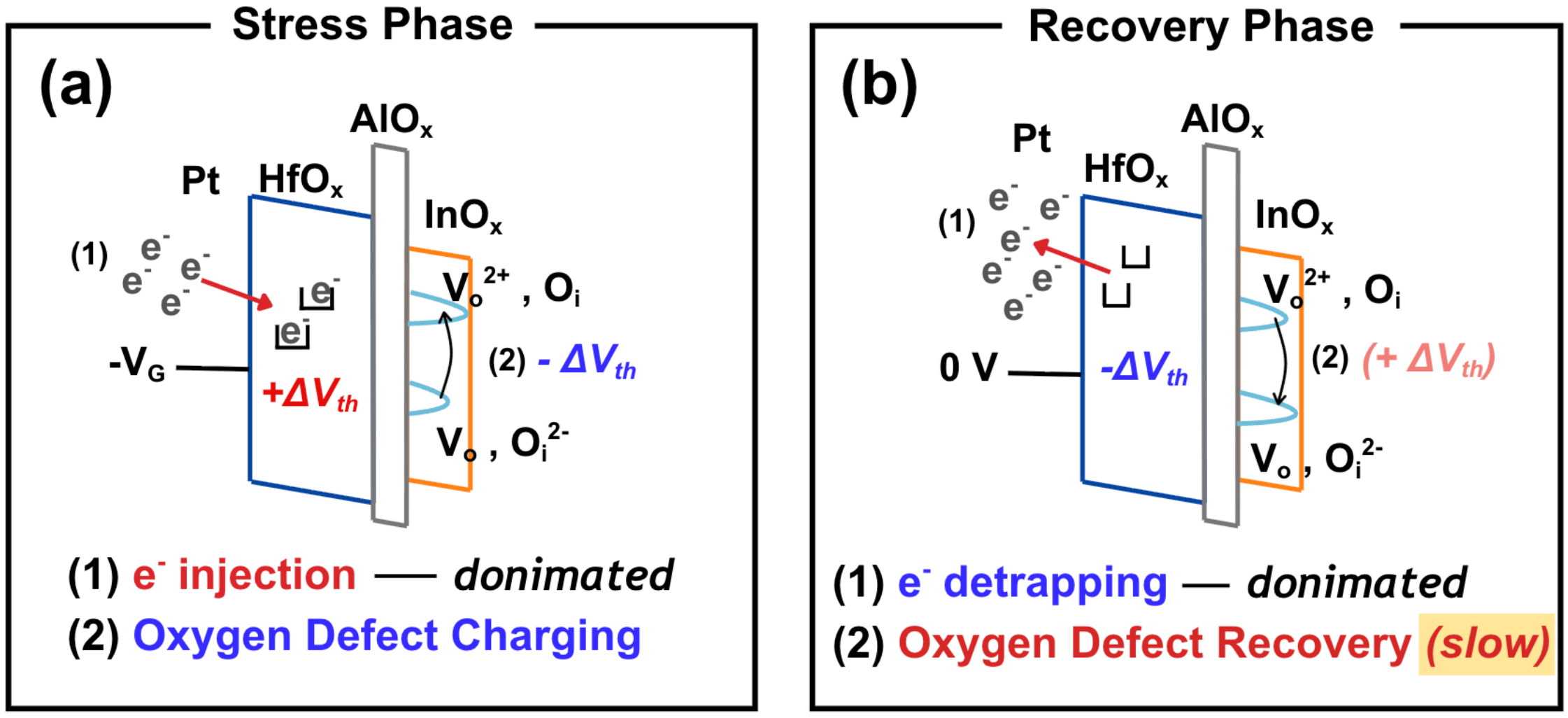

**Figure 11.** Possible instability mechanism of $InO_x$-FET with $HfO_x/AlO_x$ gate dielectric during (a) NBTI and (b) recovery.

Figures 11(a) and 11(b) summarize the proposed mechanisms during NBTI and recovery in the $HfO_x/AlO_x$-gated $InO_x$ FETs. During NBTI, gate-injected electron trapping in $HfO_x/AlO_x$ produces a positive contribution to $\Delta V_{th}$, while channel defect charging produces an opposing negative contribution. The measured positive shift indicates that the former contribution dominates under the investigated stress conditions. During recovery at $V_g$ = 0 V, the positive $\Delta V_{th}$ is recovered through electron detrapping, while a more persistent negative $\Delta V_{th}$ remains. Their different recovery kinetics can account for the observed negative overshoot. Metastable oxygen-related configurations, including peroxide defects, are possible microscopic candidates for the slowly recovering component.

## 4. Conclusion

The NBTI and recovery characteristics of ultrathin $InO_x$ FETs with $SiO_2$ and $HfO_x/AlO_x$ gate dielectrics were investigated between 25 and 85 °C. Under the same channel potential conditions, the $SiO_2$-gated devices exhibited negative threshold-voltage shifts, whereas the $HfO_x/AlO_x$-gated devices showed positive shifts that increased with temperature and $HfO_x$ thickness. Subsequent recovery at zero gate bias revealed a negative recovery overshoot in the $HfO_x/AlO_x$-gated devices, with a similar residual negative shift in different $HfO_x$ thicknesses but larger at higher temperatures. The contrasting stress and recovery behaviors are attributable to a fast and recoverable positive contribution from gate-injected electron trapping in a $HfO_x/AlO_x$ dielectric and an opposing, more persistent contribution associated with channel defect charging. Metastable oxygen-related configurations, including peroxide defects, provide a possible microscopic explanation for the persistent negative component, although their specific role remains to be established. These findings indicate that a small net threshold-voltage shift during NBTI can mask competing instability contributions and the stress recovery of oxide-semiconductor FETs can be complicated based on different kinetics of multiple mechanisms.

## Acknowledgements

This work was supported by JSPS KAKENHI Grant Number JP24K17328 and JP26K01015, JST ASPIRE Grant Number JPMJAP2534, and JST-ALCA-Next Program Grant Number JPMJAN23E4.